\documentclass[letter]{aa} 

\usepackage{graphicx}
\usepackage{txfonts}
\begin{document} 

   \title{Asteroseismology of evolved stars to constrain the internal transport of angular momentum.}


   \subtitle{III. Using the rotation rates of intermediate-mass stars to test the Fuller-formalism}

   \author{J.W. den Hartogh
          \inst{1,2,3}
          \and
          P. Eggenberger \inst{4}
          \and
          S. Deheuvels \inst{5}
          }

   \institute{Konkoly Observatory, MTA CSFK, 1121, Budapest, Konkoly Thege Miklós út 15-17, Hungary\\
   \email{jacqueline.den.hartogh@csfk.mta.hu}
   \and
   Astrophysics group, Lennard-Jones Laboratories, Keele University, ST5 5BG, UK
   \and 
   NuGrid Collaboration, \url{http://www.NuGridstars.org}
   \and
   Observatoire de Gen\`{e}ve, Universit\'{e} de Gen\`{e}ve, 51 Ch. des Maillettes, 1290 Sauverny, Suisse
   \and
   IRAP, Université de Toulouse, CNRS, CNES, UPS, Toulouse, France
             }

   \date{Received 24 January 2020; accepted 10 February 2020}

 
  \abstract
   {The internal characteristics of stars, such as their core rotation rates, are obtained via asteroseismic observations. A comparison of core rotation rates found in this way with core rotation rates as predicted by stellar evolution models demonstrate a large discrepancy. This means that there must be a process of angular momentum transport missing in the current theory of stellar evolution. A new formalism was recently proposed to fill in for this missing process, which has the Tayler instability as its starting point \citep[by][hereafter referred to as `Fuller-formalism']{2019Fuller}.}
   {We investigate the effect of the Fuller-formalism on the internal rotation of stellar models with an initial mass of 2.5 M$_{\odot}$. }
   {Stellar evolution models, including the Fuller-formalism, of intermediate-mass stars were calculated to make a comparison between asteroseismically obtained core rotation rates in the core He burning phase and in the white dwarf phase.}
   {Our main results show that models including the Fuller-formalism can match the core rotation rates obtained for the core He burning phases. However, these models are unable to match the rotation rates obtained for white dwarfs. When we exclude the Fuller-formalism at the end of the core He burning phase, the white dwarf rotation rates of the models match the observed rates. }
   {We conclude that in the present form, the Fuller-formalism cannot be the sole solution for the missing process of angular momentum transport in intermediate-mass stars.}


   \maketitle
%
\section{Introduction}
Recent developments in asteroseismology have led to a wealth of information on the internal properties of mainly low- and intermediate-mass stars \citep[recently reviewed in ][]{Aerts2018}. In particular, the analysis of mixed modes have been used to infer core rotation rates in post-main sequence stars \citep{2012beck,2012deheuvels,2012Mosser,Deheuvels2014,Deheuvels2015,2016DiMauro, 2017deheuvels,2018Gehan,2018DiMauro,2019Tayar}. A large discrepancy exists between the predicted values of stellar evolution models and observed values  \citep{2012eggenberger,2013ceillier,Marques2013,2014cantiello}, which confirms the findings of earlier works \citep{pin89,cha95,egg05,2008Suijs,Denissenkov2010}. \\
This discrepancy between predicted and observed core rotation rates shows that an additional, dominant process that transports angular momentum (AM) from the core to the outer layers is active in stars, but missing from stellar evolution codes. There is still no consensus on the physics behind this discrepancy. Several physical processes are currently being investigated. One is presented in the work by \citet{2015kissin}, who argues that magnetic torques leads to solid-body rotation in radiative regions, and differential rotation in convective regions. Another physical process being investigated is transport of AM by internal gravity waves; see \citet{2013Rogers}, and an extensive overview in \citet{Aerts2018}. Magnetised winds have been tested by \citet{2018tayar,2019Tayar}. Recently, a re-investigation of the Tayler instability was published by \citet{2019Fuller} (hereafter referred to as the `Fuller-formalism'). There is no consensus on which of these options, or whether any of them at all, are actually present in stars and acts as a dominant process in terms of AM transport. \\
In this study, we focus on testing the Fuller-formalism on a small grid of intermediate-mass stars. Previous tests of this formalism highlighted that it indeed transports AM and is dominant over the meridional and shear instabilities \citet{2019Eggenberger,2019cEggenberger}. These tests, which have only concerned low-mass stars, show that the Fuller-formalism in its current form is not the full solution to the missing process of AM transport. In particular, \citet{2019cEggenberger} shows that the Fuller-formalism has trouble reproducing the core rotation rates of subgiant stars and argues that different values of free parameter $\alpha$ are needed to match consecutive stellar evolutionary phases.\\
In this study, we focus on the comparison of the Fuller-formalism with the asteroseismic observations of intermediate-mass stars \citep{Deheuvels2015}. The seven core He burning stars in that study all show weak radial differential rotation and slow core rotation rates. Recently, these findings were confirmed to be a general characteristic of core He burning stars with a mass between 2 to 3 M$_{\odot}$ and not merely an observational bias, by \citealt{2019Tayar}. This particular initial mass range is important as it is known to be the site for the main component of the slow neutron capture process and, thus, responsible for about 50\% of all elements heavier than iron \citep{falk_ARAA,Kappeler2011,2014PASA}. Finally, many of the known white dwarfs whose rotational periods have been measured \citep{2015kawaler,Hermes2017} are believed to have progenitors in the same mass range of 2 to 3 M$_{\odot}$. In the following section we introduce the physics of our stellar evolution models and our results are shown in Section \ref{sec:results1}. Our conclusions are summarised in Section \ref{sec:concl}.

\section{Stellar evolution models}
The stellar evolution models are generated with MESA revision 8845 \citet{2011_MESA_1,2013_MESA_2,Paxton2015}. The initial settings are the same as in \cite{Jacqueline1,Jacqueline2}, as is the method we used to get through the convergence issues that can occur at the end of the AGB phase. The implementation of the Fuller-formalism is the same as in \citet{2019Fuller}, so that only the formalism's effect on the transport of angular momentum is included in the stellar evolution models and not its effect on the mixing of chemical elements. No other processes linked to rotation are included in the calculations in our approach, which is also in agreement with \citet{2019Fuller}.\\ 
In \citet{2002spruit}, the magnetic torque directly follows from the physics of the Tayler instability itself as it interacts with differential rotation. In \citet{2019Fuller}, the torque follows from a turbulence formalism that focusses on energy dissipation. The resulting energy dissipation in the Fuller-formalism is then lower than in the TS-dynamo, leading to stronger magnetic fields and Maxwell stresses. As a result, the Fuller-formalism is more efficient in transporting AM than the TS-dynamo. The equations for the transport of AM in the Fuller-formalism are:
\begin{align}
&\nu_{\mathrm{F}}=\alpha^3 r^2 \Omega \left( \frac{\Omega}{N_{\mathrm{eff}}}  \right)^2, \\
&N^2_{\mathrm{eff}}=\frac{\eta}{K}N^2_{\mathrm{T}}+N^2_{\mu},\\
&q_{\mathrm{min}}=\alpha^{-3} \left( \frac{N_{\mathrm{eff}}}{\Omega} \right)^{5/2} \left( \frac{\eta}{r^2\Omega} \right)^{3/4},
\end{align} 
with $\alpha$ being a dimensionless factor (discussed below), $\eta$ and $K$ the magnetic and thermal diffusivities, $N_{\mathrm{eff}}$ the effective Brunt-V\"ais\"al\"a frequency, and $q_{\mathrm{min}}$ the new minimum threshold for shear (q=-$\frac{\partial\mathrm{ln} \Omega}{\partial \mathrm{ln} r}$). The Fuller-formalism is only active and transporting AM when the shear is larger than $q_{\mathrm{min}}$.\\

\begin{equation}
\left( \frac{\partial \Omega}{\partial t}\right)_m=\frac{1}{j}\left( \frac{\partial}{\partial m}\right)_t \left[ (4\pi r^2 \rho)^2j D_{\rm{am}} \left( \frac{\partial \Omega}{\partial m}\right) \right]-\frac{2\Omega}{r}\left(\frac{\partial r}{\partial t}\right)_m\left( \frac{1}{2}\frac{d\rm{ln}\mathit{j}}{d\rm{ln}\mathit{r}}\right), 
\end{equation}
where $\Omega$ is the angular velocity, $j$ the specific angular momentum, and the total diffusion coefficient for AM transport is $D_{\rm{am}}$. Following \citet{2019Fuller}, 1) the only process included in the AM transport is the Fuller-formalism and, thus, $D_{\rm{am}}$=$\nu_{\rm{F}}$ ; and 2) there is no mixing of chemical elements due to rotation included in the models presented here.\\
The small set of models calculated for this work have an initial mass of 2.5 M$_{\odot}$, an initial metallicity of Z=0.01, and an initial rotation rate of 50 km/s; this follows a direct continuation of \cite{Jacqueline1}. The Fuller-formalism has only one free parameter\footnote{There are a few more free parameters, but those only concern the smoothing of $\nu_{\rm{F}}$. Their values are the same as in \citealt{2019Fuller}.}: $\alpha$, which we vary in order to match the core rotation rates of the seven stars analysed in \citet{Deheuvels2015}.\\
These seven core He burning stars were chosen because\ both the core and surface rotation rate were known, which allows for an improved constraining of the initial parameters. The small size of this sample made it difficult to determine whether it was sufficiently representative of  intermediate-mass core He burning stars, especially as their core rotation rates are used to place them among the slowest rotating core He burning stars, see Fig. 4 in \citealt{Aerts2018}. Recently, 99 new core rotation rates for core He burning stars were published by \citet{2019Tayar}, in which most stars have a mass of 2-3 solar masses. The core rotation rates of the seven stars of \citet{Deheuvels2015} fit well in the new data set and they are, thus, representative values. We include all the known white dwarfs rotation rates, as presented in \citet{2015kawaler} and \citet{Hermes2017}.

\section{Results}
\label{sec:results1}
\begin{figure}[t]
    \centering
    \includegraphics[width=\linewidth]{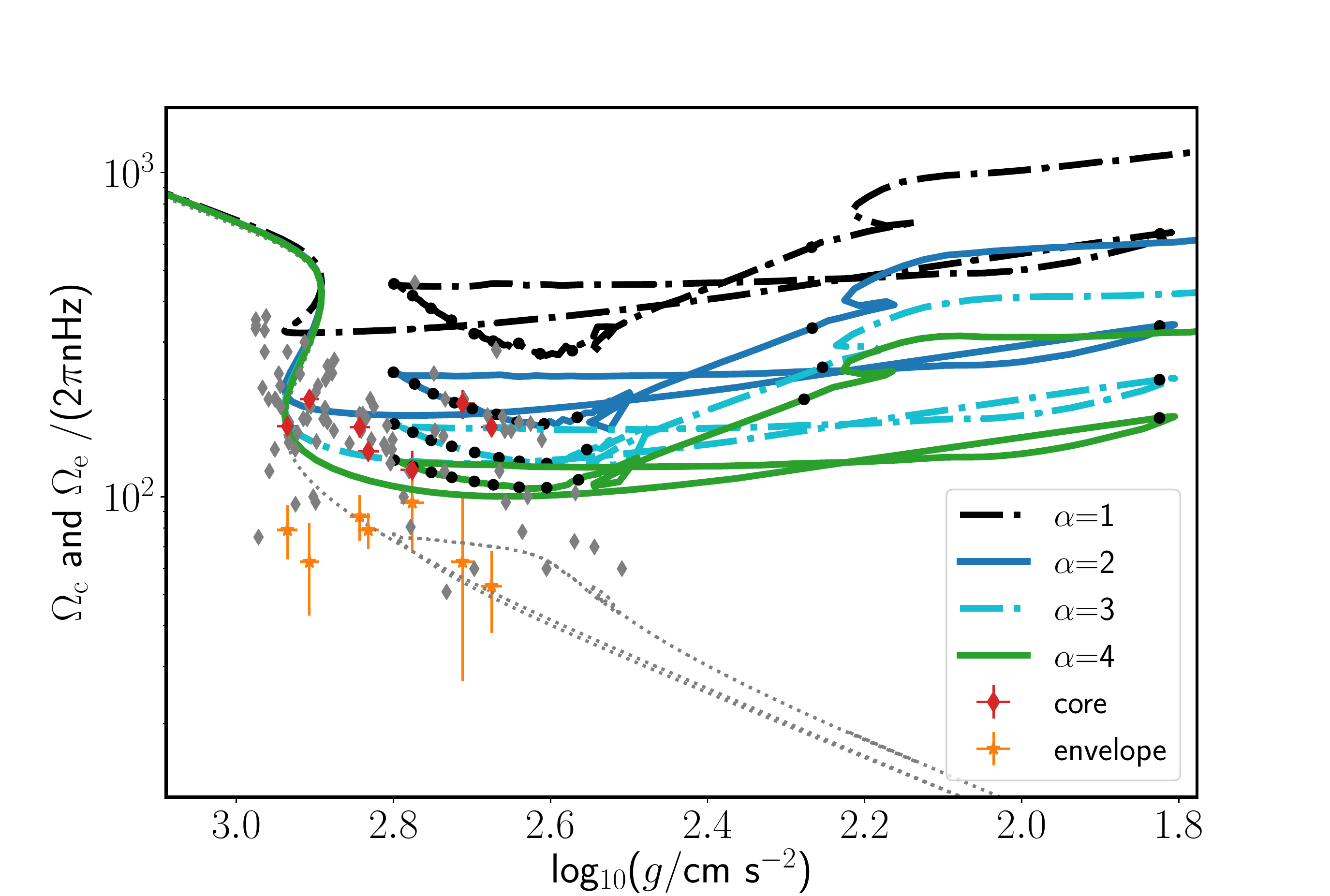}
    \includegraphics[width=\linewidth]{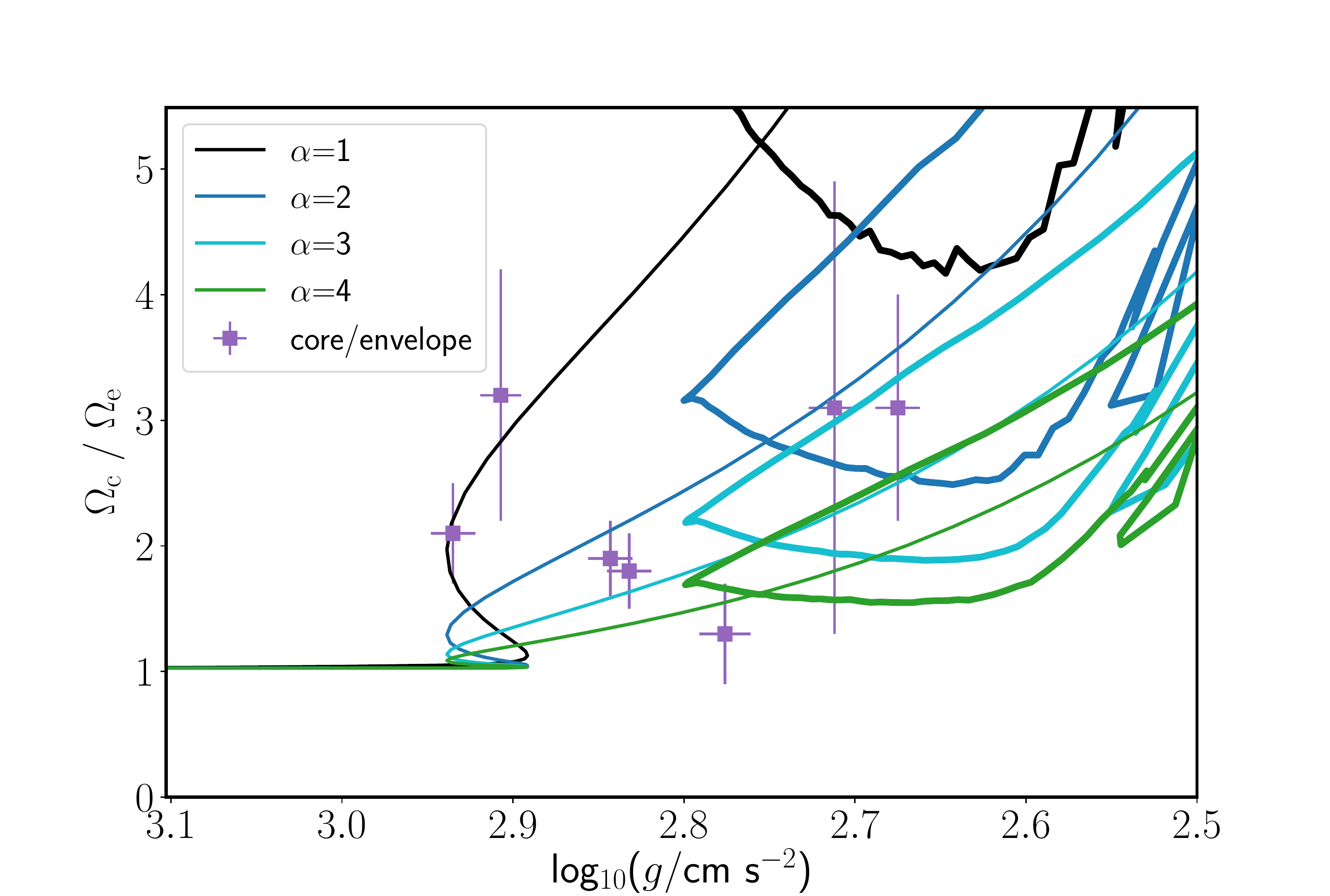}
    \caption{Comparison of models with different alphas to all observed core rotation rates (top panel) within 2-3M$_{\odot}$ range. Highlighted in colour are the core He burning stars of \citet{Deheuvels2015}, showing the core (red diamonds) and surface (orange stars) values in the top panel and the ratio of the two values (purple squares) in the bottom panel. The core rotation rates of \citealt{2012Mosser} and \citealt{2019Tayar} are shown in grey. Values of $\alpha$ > 1 are needed to reach the observed range of core rotation rates and degree of differential rotation. One surface rotation rate is plotted (dotted line) in the left panel. The surface rotation rates of the other models are identical to the one plotted and, therefore, they are not included in this figure.}
    \label{fig:omegas}
\end{figure}

The evolution of the core rotation rates of the 2.5 M$_{\odot}$ models, including the Fuller-formalism, is shown in the top panel of Fig. \ref{fig:omegas}. The core rotation rates correspond to the mean value determined in the g-mode cavity \citep[following][]{2013Goupil} and are plotted against the surface gravity. Black dots are added to indicate the core He burning phase, where interval between dots accounts for 10\% of the total duration of the core He burning phase. The first dot (located around log$_{10}(g$/cm s$^{-2})\simeq$ 1.8) indicates the start of the core He burning phase, while the final black dot (located around log$_{10}(g$/cm s$^{-2})\simeq$2.3) indicates the end of the core He burning phase. These markers show that the stellar models spend the majority of the phase in the part of the parameter space of $\Omega_{\rm{c}}$ , where the observed rotation rates are located. 
A slightly lower initial mass is needed in the stellar evolution models to reach all the corresponding log$_{10}g$ values.\\
In order to reach the values by \citet{Deheuvels2015}, we had to include a value for $\alpha$ of two to four.
\noindent This increase enhances the transport of angular momentum, leading to lower core rotation rates. The grey data points in the top panel are the core rotation rates of stars with a mass between 2 and 3 M$_{\odot}$ of \citet{2012Mosser} and \citet{2019Tayar}, which cover a similar space in the figure. This indicates that these values for $\alpha$ are needed for all stars in this mass range, possibly even higher values for the data points just below $\Omega_{\rm{c}}=$10$^2$ nHz.\\
The increase of $\alpha$ also results in a decrease of the differential rotation, which improves the match of the models to the ratio of core to surface rotation rate, as shown in the bottom panel of Fig. \ref{fig:omegas}. The need for these higher values of $\alpha$ is in contrast to \citet{2019Fuller}, who suggests that $\alpha$ should be around 1. \\
These results are different in several aspects from the ones obtained by \citet{Jacqueline1}, using a constant additional, artificial viscosity to match the seven stars from \citet{Deheuvels2015}. The models with the Fuller-formalism experience solid body rotation until a log$_{10}(g$/cm s$^{-2})\simeq$ 2.9 is reached, while the models in \citet{Jacqueline1} diverge from solid body rotation at the end of the main sequence, at log$_{10}(g$/cm s$^{-2})\simeq$ 4.0 (see their Fig. A.2 and A.3). The elongation of the solid body phase is also visible in Fig. 3 of \citet{2019Fuller}. Another difference is, of course, that the viscosity due to the Fuller-formalism is not a constant. 
When comparing our results to \citet{2019Fuller}, and specifically their Fig. 4, some differences can be identified. In particular, in Fig. 4 of \citet{2019Fuller}, the range of clump rotational periods is matched with $\alpha$ of 0.5 to 2, while we need a larger range for $\alpha$ to match all observed values. This can be explained by two points: the models presented here are calculated with a higher initial mass than in \citealt{2019Fuller}: namely, here it is 2.5 M$_{\odot}$, there it is 1.5 M$_{\odot}$. And, as shown in Fig. 6 of \citealt{2019Fuller}, their models (all using $\alpha$=1) with initial mass of 2.2 and 3.0 M$_{\odot}$ range show lower periods during their clump phase when compared to their models with an initial mass of 1.2 and 2.0 M$_{\odot}$ (respectively, 40 days and 100-200 days). These lower periods would not match the observed range as included in their Fig. 4, which is in agreement with our results.  

\begin{figure}
    \centering
    \includegraphics[width=\linewidth]{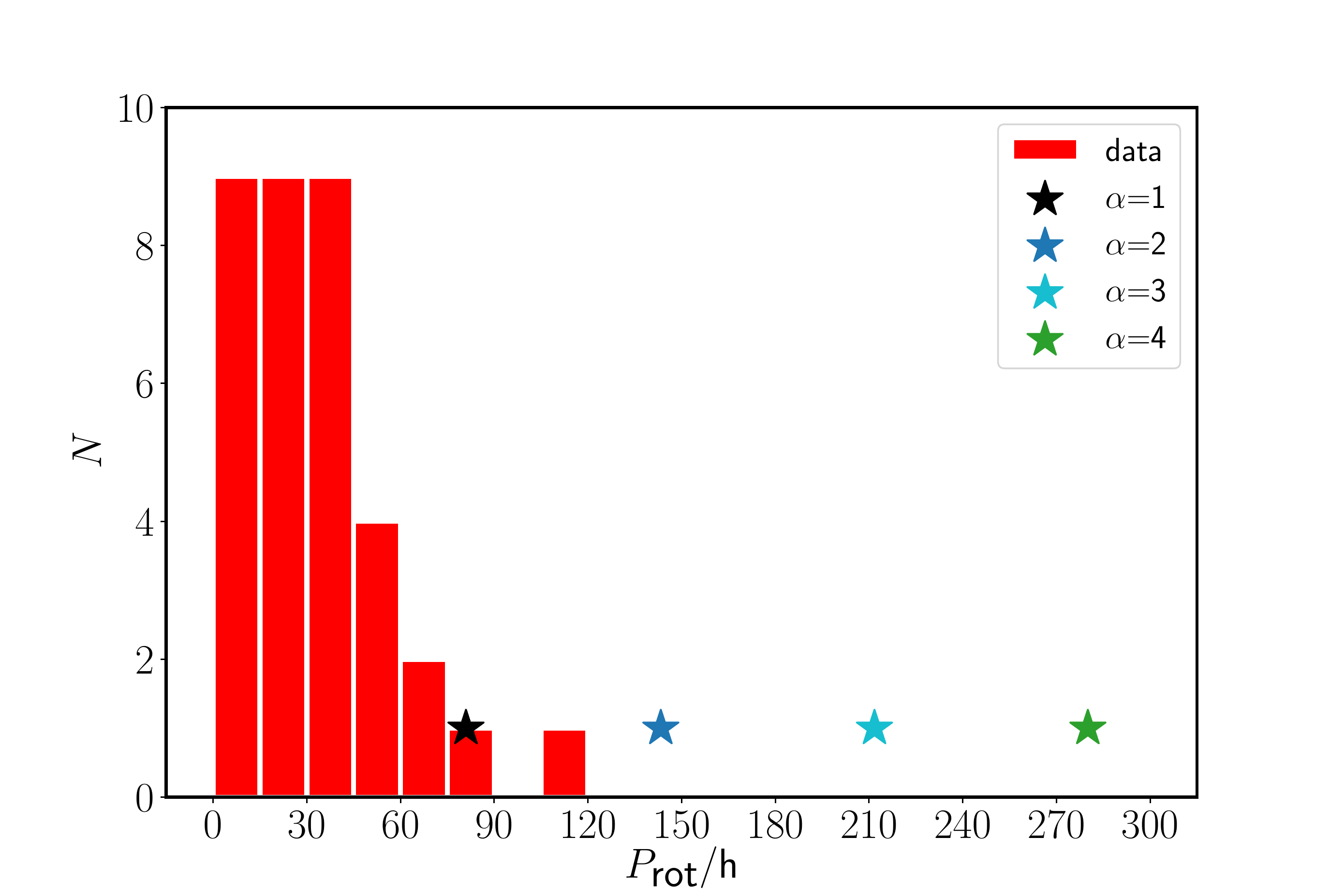}
    \caption{Comparison of observed white dwarf rotational periods and white dwarf periods from models presented here (star symbols). Only the period of the $\alpha$=1 model falls within the observed range of periods, but this model does not reach the observed core rotation rates in the core He burning phase. All other stellar evolution models reach the DAV temperature range with a rotational period higher than those observed.}
    \label{fig:wdspinshist}
\end{figure}

The stellar evolution models are continued into the white dwarf phase by using the same mass loss parameters as in \citet{Jacqueline1,Jacqueline2}. The comparison between all known white dwarf rotational periods \citep{2015kawaler,Hermes2017} and the rotational periods of the models when reaching the DAV temperature range are shown in Fig. \ref{fig:wdspinshist}. This histogram shows that most observed white dwarfs have rotational periods of two days or less, with the longest observed rotational period being 109.1 hours. We note that the observation method of \citet{Hermes2017} is sensitive to larger rotation periods up to 15 days, but the longest period they found was about 4.5 days.\\
The white dwarf rotational periods of our models start at higher values. Specifically, the stellar models that match the rotation rates of \citet{Deheuvels2015}, with $\alpha$=2-4, lead to white dwarf rotational periods that are all higher than the observed periods.\\
Similarly to the case with our results of the core He burning phase, we compare this result to the results presented in \citealt{2019Fuller}. In their Fig. 7, they compare their white dwarf rotational periods (all with $\alpha$=1) to the observed values. Our results agree that these models match the observations, however, as we explained before: the models with $\alpha$=1 do not match the low rotation rates of the clump stars analysed by \citet{Deheuvels2015}, and other clump stars in the same initial mass range of 2-3 M$_{\odot}$ by \citet{2012Mosser} and \citet{2019Tayar}.\\
\begin{figure}
    \centering
    \includegraphics[width=\linewidth]{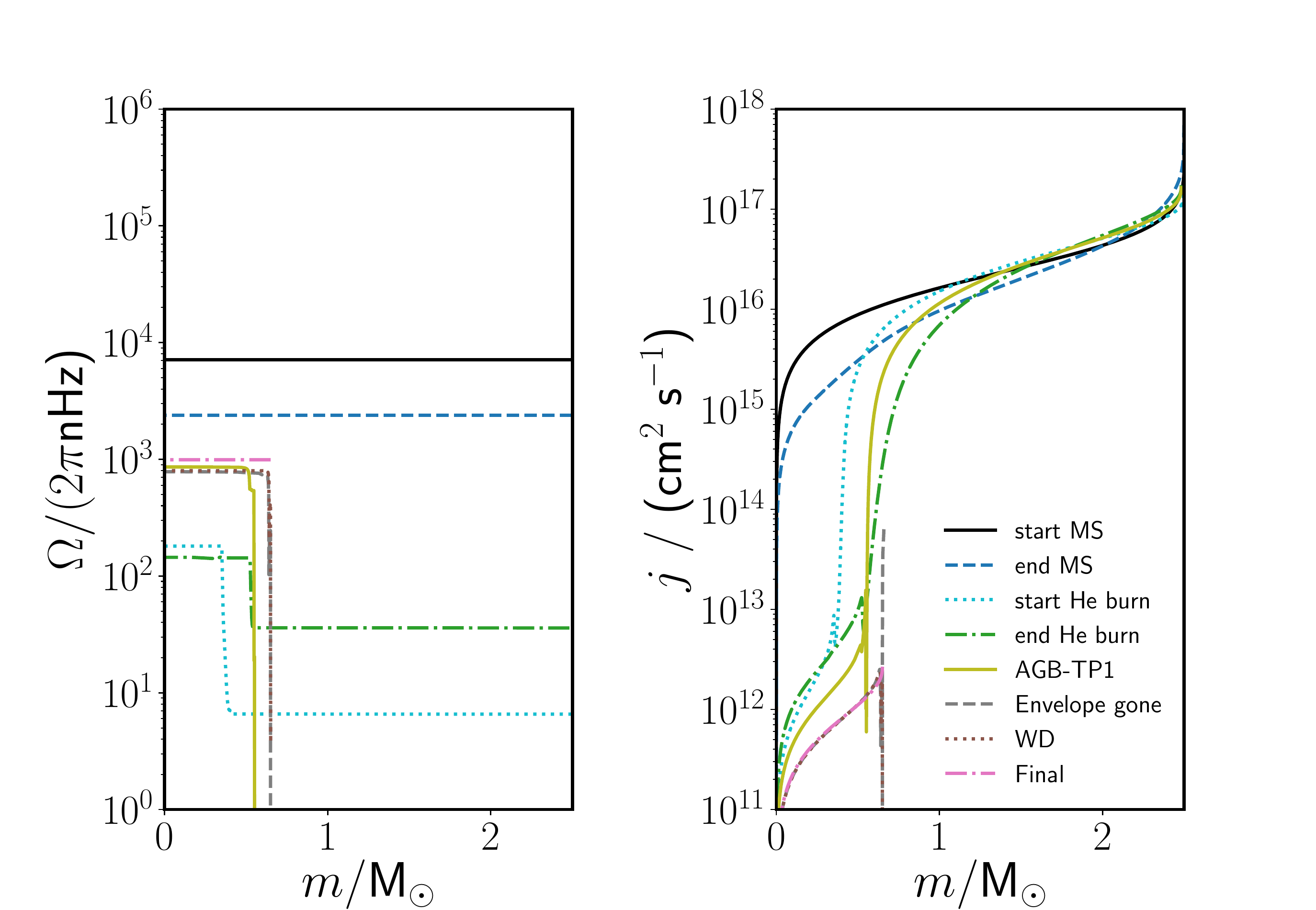}
    \caption{Profiles of $\Omega$ and $j$ for stellar evolutionary phases taken from the model with $\alpha$=4. The eight phases are the start and end of the main sequence (respectively `start MS' and `end MS'), the start and end of the core He burning phases (`start He burn', and `end He burn'), the first thermal pulse in the AGB phase (`AGB TP1'), the end of the AGB phase (`Envelope gone'), the start of the white dwarf phase (`WD'), and the final profile in the calculation (`Final'), which is located lower on the white dwarf cooling track than the DAV phase. The figures show that there is AM transport after the core He burning phases.}
    \label{fig:profiles}
\end{figure}
We further investigate the post-core He burning evolution of the core rotation rates by analysing the $\Omega$ and specific angular momentum $j$ profiles throughout the stellar evolution. These profiles are shown in Fig. \ref{fig:profiles}, from the start of the main sequence `start MS' until the final model number in the calculation 'Final', which is after the calculation has gone through the DAV temperature range. The transport of AM due to the Fuller-formalism leads to the decrease of a factor of 15 in the core rotation rate (left panel of Fig. \ref{fig:profiles}) between the end of the main sequence and the start of core He burning. This decrease is also visible the profiles of $j$. In contrast to the findings of \citet{2014cantiello, Jacqueline1} however, there is also AM transport happening after the core He burning phase. This decrease of $j$ in the core of about one order of magnitude reduces the core rotation rate and, thus, increases the rotational periods we calculate for the white dwarfs. We therefore calculate stellar evolutionary models that include the Fuller-formalism until the end of the core He burning phase and exclude it going forward. The resulting white dwarf rotational periods for stellar evolution models with $\alpha$=1 and $\alpha$=4 are shown in Table \ref{tab:myWDspins} (as `1*' and `4*', respectively). These new periods are 35\% and 25\%. respectively, of the original period calculated with the inclusion of the Fuller-formalism throughout the evolution. This decrease is large enough for the new periods to fall within the observed range of white dwarf rotational periods. 
\begin{table}
    \caption{Rotational periods of our white dwarf models, calculated in the DAV temperature range. The first four models are also shown in Fig. \ref{fig:wdspinshist}, the final two models include the Fuller-formalism from the main sequence to the end of core He burning, and exclude the formalism from then onward. The latter two periods are both within the observed range.}
    \centering
    \begin{tabular}{c|c}
       $\alpha$  & P$_{\rm{DAV}}$ (hours)\\
       \hline
         1 & 81.0\\
         2 & 143\\
         3 & 212\\
         4 & 280\\
         \hline
         1* & 28.4  \\
         4* & 70.0\\
    \end{tabular}

    \label{tab:myWDspins}
\end{table}

 
\section{Conclusions}
\label{sec:concl}
In this letter, we test the Fuller-formalism in stellar evolutionary models with an initial mass of 2.5 M$_{\odot}$. We investigate whether this formalism can lead to core rotation rates that match the observational range of rotation rates during the core He burning phase and the DAV temperature range. Our findings are:\\
\begin{itemize}
    \item When including the Fuller-formalism in our stellar models, we need an $\alpha$ value of at least 2 to reach the core rotation rates of the intermediate stars analysed by \citet{Deheuvels2015}, while lower values are needed to match core rotation rates of low mass star \citep{2019cEggenberger};
    \item After continuing with these stellar models with $\alpha$=2-4 to the white dwarf phase, we found that the resulting white dwarf rotational periods are in the range of 81-280 hours;
    \item The models that match the range of core He burning stars of \citet{Deheuvels2015} do not match the observed white dwarf rotation rates, and vice versa; 
    \item The range of observed white dwarf rotation rates is matched when the Fuller-formalism is only included from the start of the main-sequence to the end of core He burning.
\end{itemize}
Therefore, it is important to use caution when adopting the Fuller-formalism in stellar evolution models. The formalism, based on \citealt{2019Eggenberger,2019cEggenberger} and this work, is not likely to be the sole missing piece in the process of AM transport, but could still could be be part of the solution. It is possible that a physical reason behind the exclusion of the Fuller-formalism at the end of core He burning may be found, which would lead to a more favourable judgement of the formalism when included in stellar evolution models of intermediate-mass stars.

\begin{acknowledgements}
We thank Henk Spruit for useful discussions. This work has been supported by the European Research Council (ERC-2016-CO Grant 724560), and the EU COST Action CA16117 (ChETEC). PE acknowledges support by the European Research Council (ERC grant agreement No 833925). SD acknowledges support from the project BEAMING ANR-18-CE31- 0001 of the French National Research Agency (ANR).  
\end{acknowledgements}

%
%
\bibliographystyle{aa} 
\bibliography{references.bib} 

\end{document}